\documentclass[aps, twocolumn,showpacs,superscriptaddress]{revtex4}

\usepackage{color}
\usepackage{amsmath}
\usepackage{amssymb}
\usepackage{amscd}
\usepackage{wasysym}
\usepackage[ansinew]{inputenc}
\usepackage[T1]{fontenc}
\usepackage{ae,aecompl}
\usepackage{hyperref}

\usepackage[dvips]{graphicx}

\newcommand{\dirk}[1]{#1}

\newcommand{\be}{\begin{equation}}
\newcommand{\ee}{\end{equation}}
\newcommand{\bea}{\begin{eqnarray}}
\newcommand{\eea}{\end{eqnarray}}
\newcommand{\nn}{\nonumber}

\begin{document}
\bibliographystyle{apsrev}

\title{Nonlocal failures in complex supply networks \\ by single link \emph{additions}}

\begin{abstract}
How do local topological changes affect the global operation and stability 
of complex supply networks? Studying supply networks on various levels 
of abstraction, we demonstrate that and how adding new links may not 
only promote but also degrade stable operation of a network. Intriguingly, 
the resulting overloads may emerge \dirk{remotely} from where such a link is added, 
thus resulting in nonlocal failure. We link this counter-intuitive 
phenomenon to Braess' paradox originally discovered in traffic networks. 
We use elementary network topologies to explain its underlying mechanism  
for different types of supply networks and find that it generically occurs
across these systems. As an important consequence, upgrading supply 
networks such as communication networks, biological supply networks or 
power grids requires particular care  because even \textit{adding} only 
single connections may destabilize normal network operation and induce 
\dirk{disturbances remotely from the location of structural change
and even global cascades of failures.}
\end{abstract}

\author{Dirk Witthaut}
\affiliation{Network Dynamics Group, Max 
Planck Institute for Dynamics and 
Self-Organization (MPI DS), D--37077 G\"ottingen, Germany}

\author{Marc Timme}
\affiliation{Network Dynamics Group, Max 
Planck Institute for Dynamics and 
Self-Organization (MPI DS), D--37077 G\"ottingen, Germany}
\affiliation{Faculty of Physics, University of G\"ottingen, 
D--37077 G\"ottingen, Germany}

\maketitle


\section{Introduction}

Stable operation of complex supply networks underlies the proper 
function of a broad range of biological and technical systems. For
instance in plants, supply networks provide nutritions and water to cells in
leaves and other plant parts \cite{Kati10}; the world wide web relies on
stable information distribution \cite{Albe00}; and electric power 
grids operate only if electricity demand matches supply at every point 
of the grid \cite{Prab94,12powergrid}.
Whereas supply networks in
biological systems are created during development and may be fixed after,
e.g., leave damages,  our technical infrastructure has to be constantly
modernized and extended to meet technical developments and the future demand.
One particular important example is the drastic change of electric power 
supply in the upcoming decades which provides an extraordinary challenge for 
the operation of future power grids \cite{Pepe05,Carr06,Blaa06}. 
New transmission lines have to be build in order to transport electric energy 
generated by wind turbines and other renewable energy sources to 
consumers at remote locations  \cite{Marr08}. 
Due to the importance of a stable power supply and the enormous expenses 
of new transmission lines, a careful planning of the optimal future network 
topology is inevitable. This planning has to pay special regard to the 
\emph{collective} dynamics of the complete network -- an isolated 
stability analysis of single elements or local subnetworks is not sufficient
\dirk{
\cite{Mott02,Scha06,Kiss07,Simo08,Nish10,Buld10,Vesp11,Pars11,Menc13}.
}
A striking example highlighting the importance of collective effects in real
supply networks is the power outage in western Europe on 4 November
2006. Here, the manual disconnection of one double-circuit power line in 
Northern Germany triggered power outages in many European countries 
up to Spain. A main reason for this fatal misplanning was the inobservance
of global effects beyond the local grid and the lack of communication 
between the different grid operators \cite{UCTE07}.

In this article, we analyze the collective response of supply networks
to the \emph{addition} of new links and clarify a surprising, 
yet very general effect: We show that whereas additional links stabilize 
the operation of the network on average, specific potentially new links 
\emph{decrease} the total network capacity and may deteriorate or even 
destroy network functionality. 
A similar effect in traffic flow was described and explained 
by Braess in a game theoretical framework \cite{Brae68,Brae05} 
and later confirmed experimentally and numerically in different 
systems \cite{Cohe91,Roug02,Blum07,Youn08,Vali10}. Related 
phenomena were already discussed much earlier in \cite{Ward52,Beck56}.
We explicate this counter-intuitive effect, referred to as Braess' paradox, for a range of model supply networks: an abstract messaging model, a flow model resembling Kirchhoff's laws
for DC networks, a static power flow model and a dynamic oscillator model of AC power grids \cite{Fila08,12powergrid}. Interestingly, adding certain links may not only cause cascading failures but also immediate overloads in parts of the network that are remote from the location of such links, thus indicating a non-local impact of specific link addition.

This article is structured as follows: After providing an intuitive explanation and a mathematical analysis of Braess' paradox for elementary model networks in Sec.~\ref{sec:basics}, we study
different aspects of this effect in large complex networks.
In Sec.~\ref{sec:cascade} we show how the addition of new links can lead 
to a major breakdown of a supply network due to a cascading failure triggered 
by Braess' paradox. 
In complex networks, topological changes such as link addition and removal  
can cause severe nonlocal effects (Sec.~\ref{sec:nonlocal}).
Finally we study in Sec.~\ref{sec:complex} which networks are susceptible to
suffer Braess' paradox and which factors facilitate it. The first identification of Braess' paradox in oscillator networks has been briefly reported recently \cite{12braess}.

\section{Braess' paradox in elementary network models}
\label{sec:basics}

We first reveal the mechanisms underlying Braess' paradox for small systems
that allow a detailed analytic description. We demonstrate the importance
of Braess' paradox in supply networks for four model classes:
an abstract messaging model introduced by Motter and Lai \cite{Mott02},
a DC power flow model, an AC power flow model and an oscillator 
model for power grid dynamics introduced by Filatrella \textit{et al.}
\cite{Fila08} for simple grids and extended to complex networks 
by Rohden \textit{et al.} \cite{12powergrid}.
A detailed description of these models is provided in appendix \ref{sec:appendix_models}.

\subsection{Braess' paradox in messaging networks}
\label{sec:braess-motlai}

A simple model for supply networks was introduced by Motter and Lai \cite{Mott02}
to study the dynamics of cascading failures. In this model every node sends one
unit of the relevant quantity, e.g. information or electric energy, to each other node
via the shortest path in the network.

\begin{figure}[tb]
\centering
\includegraphics[width=8cm, angle=0]{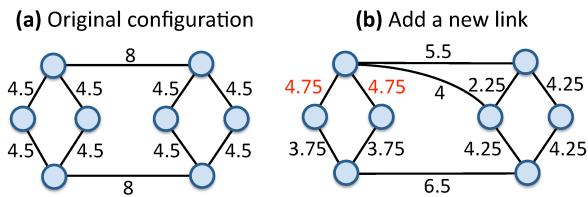}
\caption{\label{fig:bmlai1}
(color online). Braess' paradox in an elementary messaging network. In one period, 
each \dirk{node sends one message to each other node} 
along the shortest path. The numbers specify the load \dirk{$F$} of each link.
If a new link is added to the network (panel b), the load at two 
links adjacent to the new link \emph{increases} from 
\dirk{$F = 4.5$ to $F = 4.75$} (red numbers).
}
\end{figure}

A basic example for Braess' paradox in such a messaging network 
is shown in Fig.~\ref{fig:bmlai1}.  The loads 
change when a new link is added. As expected, the total load is distributed 
over more links, such that the load decreases at most links (and on average). However, there are two edges of the network, where the load \emph{increases} 
when the new edge is \emph{added}.
If the capacity of these links is limited to $K < K_c = 4.75$, they will become
overloaded and drop out of service. This local failure may cause a 
breakdown of the complete network by a cascade of failures  -- 
an effect which will be studied in detail in Sec.~\ref{sec:cascade}.
This simple example already shows a basic mechanism which triggers
Braess' paradox. When a new link is added, it can provide a shortcut 
for messages. This often increases the loads of some links \emph{connecting} to the added one and these overloaded links, in turn, drop out of service (see also Sec.~III).

\subsection{Braess' paradox in flow networks}
\label{sec:braess_flow}

The messaging model briefly discussed above provides an elementary, well accessible model for supply networks, but relies on rather specific assumptions. Supply
comes only in discrete units and is transported via the shortest paths, 
regardless of the capacity of the respective links. For general supply 
networks, more sophisticated models are needed.

\begin{figure}[t]
\centering
\includegraphics[width=8cm, angle=0]{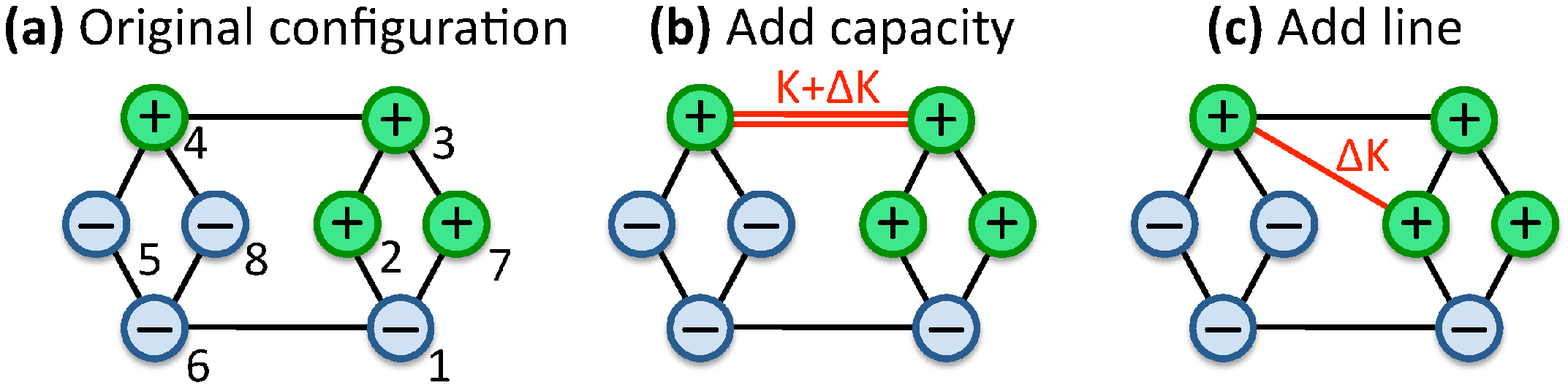}
\includegraphics[width=8cm, angle=0]{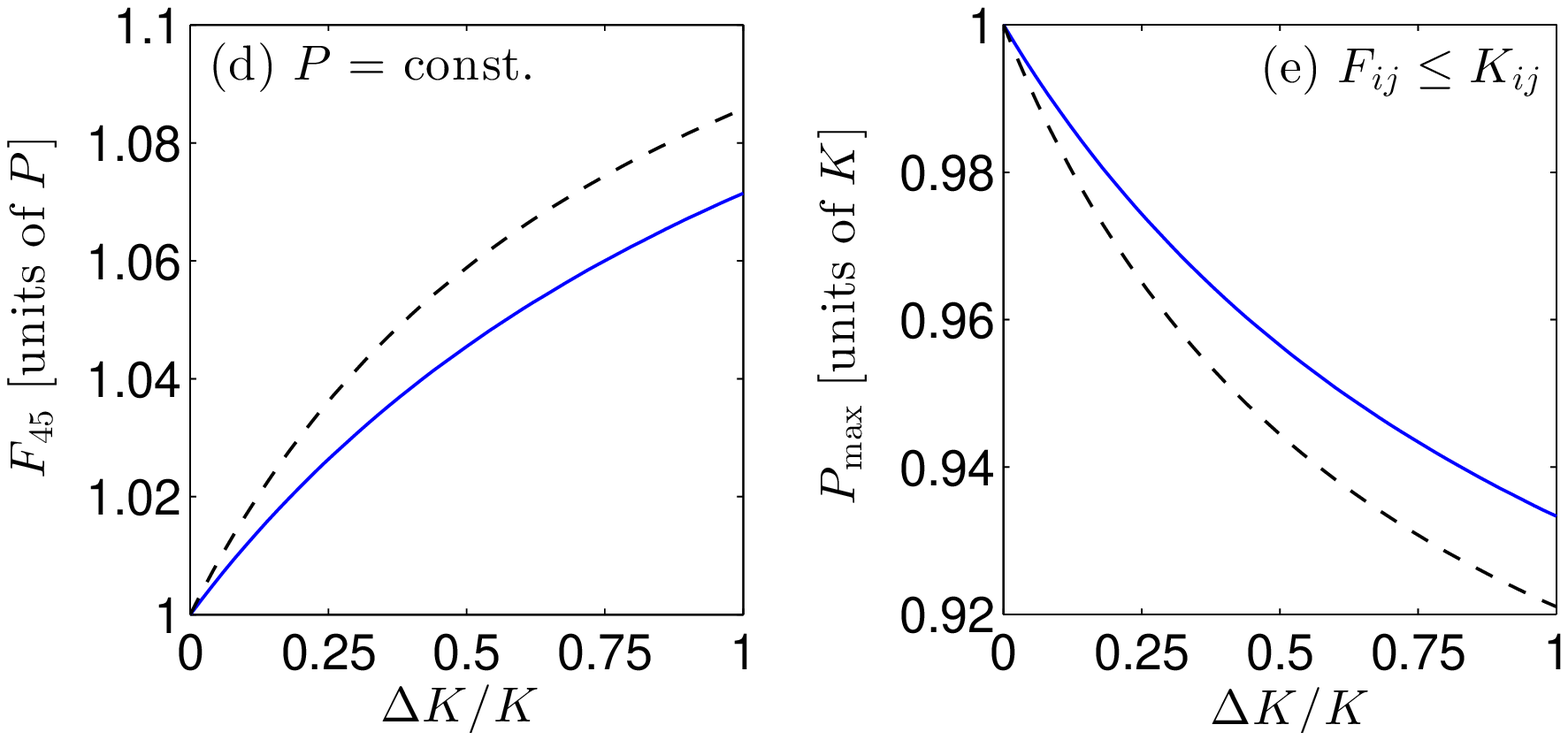}
\caption{\label{fig:bflowa1}
(color online) {Braess' paradox in a flow model.} 
(a) Topology of the network. The vertices generate/consume
the power $P_j = \pm P$. All edges have the same transmission 
capacity $K$.
(b,c) Additional transmission capacity is provided by upgrading
one link (b) or adding a new link (c).
(d) Load of the link $4\leftrightarrow5$ as a function of the 
additional capacity if $P$ is kept constant.
(e) Maximum transmittable power $P$ if the capacity of each link is
limited as $F_{ij} \le K_{ij}$.
Results are shown for additional capacity (scenario (b), ---) and
an additional link (scenario (c), - - -).
}
\end{figure} 

Now, we first consider an elementary flow model which models, for instance,
DC power grids or biological supply networks \cite{Kati10}.
The network is specified by the transmission capacity $K_{ij} > 0$ between the 
nodes $i,j \in \{ 1,\ldots, N \}$, where $N$ denotes the number of nodes in 
the network. Obviously, we have $K_{ij} = K_{ji}$ and we set $K_{ij} = 0$ if 
no link exists between nodes $i$ and $j$. Furthermore, each node of the 
network is characterized by the electric power $P_j$ it generates ($P_j > 0$) 
or consumes  ($P_j < 0$). 
We denote the flow from node $i$ to node $j$ by $F_{ij}$, which can be positive 
(power flows from $i$ to $j$) or negative (power flows from $j$ to $i$). 
The conservation of energy then reads
\be
   \sum_{j=1}^{N} F_{ij}  = P_i \qquad 
       \text{ for all } j \in \{1, \ldots ,N \}.
   \label{eqn:flowcon1}
\ee
\dirk{In general, this condition is not sufficient to 
uniquely fix the flows $F_{ij}$ (see appendix \ref{sec:flowmodel}).}
We furthermore assume that the unique steady state is determined by the 
condition that the total dissipated power
\be
    E_{\rm diss} =  \sum \nolimits^{'}_{i<j} \frac{F_{ij}^2}{2 K_{ij}}
    \label{eqn:Edissflow1}
\ee
is minimal. Here, the primed sum runs over
all existing links, i.e. all pairs $(i,j)$ with $K_{ij} \neq 0$. 
For DC power grids these two conditions then imply
Kirchhoff's circuit laws (see appendix \ref{sec:flowmodel}).

An elementary example for the occurrence of Braess' paradox in
this flow model is shown in Fig.~\ref{fig:bflowa1}. The upper panels
show the original network structure and two possible scenarios of upgrading
the grid --
either the capacity of the upper link is increased by an amout $\Delta K$ 
or another link with capacity $\Delta K$ is added. The remaining
links have a capacity $K$ and the nodes generate ($+$) or consume
($-$) the power $P$.

In the original network, a stable failure-free operation is possible as long
as $P \le P_{\rm max} = K$. On the edge of the stability region, i.e. for
$P = P_{\rm max}$, the following six links are maximally loaded:
\be
   6 \leftrightarrow 1, \quad 
   1 \leftrightarrow 2, \quad 
   1 \leftrightarrow 7, \quad 
   3 \leftrightarrow 4, \quad 
   4 \leftrightarrow 5, \quad 
   8 \leftrightarrow 8. \nn
\label{eq:maxLoadedLinks}
\ee 
If additional transmission capacity or a new link is added to the network as 
shown in Fig.~\ref{fig:bflowa1} (b,c), the load of the adjacent links
$F_{45} = F_{48}$ connecting to the new link increases as shown in 
Fig.~\ref{fig:bflowa1} (d). Hence, these links are crucial for a failure-free 
operation of the complete network. If the original network is already 
operating close to the edge of the stability region, the addition of new 
capacity or a new link can damage the links $4 \leftrightarrow 5$ and 
$4 \leftrightarrow 8$ which may then trigger a power outage in the network. 
Therefore,
the \emph{increase} of local transmission capacity $\Delta K$ 
leads to a \emph{decrease} of the maximum power $P_{\rm max}$ which
can be transmitted through the network as shown in Fig.~\ref{fig:bflowa1} (e).
We note that $P_{\rm max}$ decreases monotonically with $\Delta K$ in this
example of a linear flow networks whereas a different behaviour has been reported for traffic flow models  \cite{Nagu10}.

The power flow in the elementary networks analyzed in Fig.~\ref{fig:bflowa1} 
can be calculated analytically, which yields a closed condition for a
failure-free operation of the supply network. In the following we 
consider the first scenario, where the capacity of the upper link 
$3 \leftrightarrow 4$ is increased by an amount $\Delta K$.
The condition of flow conservation (\ref{eqn:flowcon1}) at each node gives rise to
eight linear equalities for the eight non-zero flows $F_{ij}$. As one of
the conditions is redundant, there is a one-dimensional family of solutions to
the linear equations parametrized by a real number
$\delta$,
\bea
   && (F_{16}, F_{21}, F_{32}, F_{43}, F_{54}, F_{65}) = (F_a-\delta F_b), \quad \mbox{where} \nn \\
   && \qquad F_a = P (-1, -1, 0, +1, +1, 0) \nn \\
   && \qquad F_b = P (-2, -1, -1, +2, +1, +1).
   \label{eqn:flows-ss}
\eea   
with 
\bea
     && F_{71} = F_{21}, \qquad \quad F_{37} = F_{32}, \nn \\
     && F_{84} = F_{54} \quad \mbox{and} \quad F_{68} = F_{65}.
\eea
due to the symmetry of the network.
Minimizing the total dissipation (\ref{eqn:Edissflow1}) with respect to the parameter
$\delta$ yields
\be
   \delta = \frac{\Delta K}{8K + 6 \Delta K} \, ,
\ee
such that the load of the critical links is given by
\be
  F_{54} = F_{84} = P  \; \frac{8 K + 7 \Delta K}{8 K + 6 \Delta K} \, .
\ee
An overload occurs if the load is larger than the capacity of the link,
i.e. if $|F_{54}| > K$. Thus we find that a failure-free operation of the
supply network is only possible if 
\be
  P  \le P_{\rm max} = K \; \frac{8 K + 6 \Delta K}{8 K + 7 \Delta K} \, .
\ee
In particular, if the additional capacity $\Delta K$ increases, the 
maximum transferable power $P_{\rm max}$ decreases.

\subsection{Braess' paradox in a AC power grid}

Braess' paradox also occurs in AC power grids, which provide the
backbone of our technical infrastructure \cite{Grai94,Prab94}. 
In this section we study 
the static operation of a grid in a power flow study before turning
for a dynamic model in the following section. 
The details of the static flow model are described in appendix 
\ref{sup:ACflow}.

\begin{figure}[t]
\centering
\includegraphics[width=7cm, angle=0]{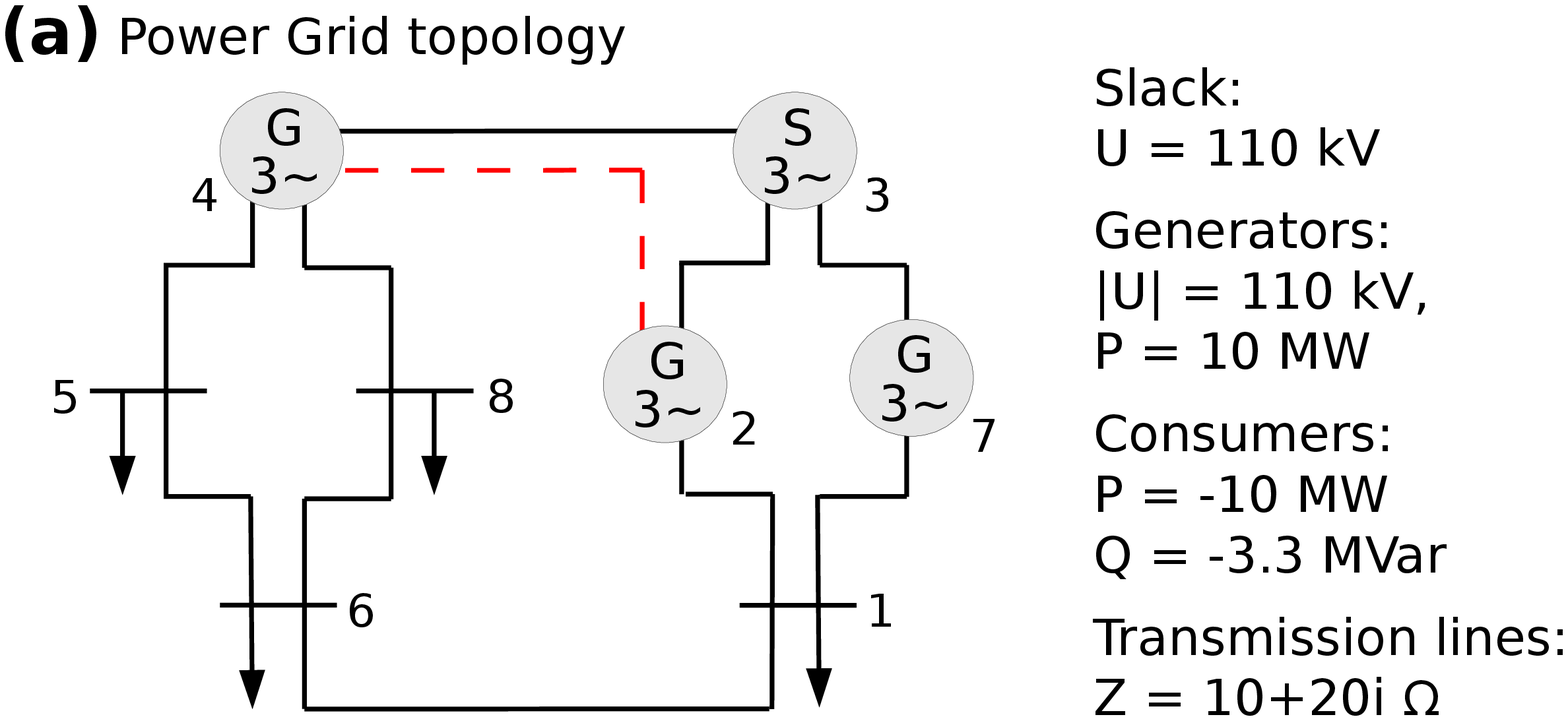}
\includegraphics[width=7cm, angle=0]{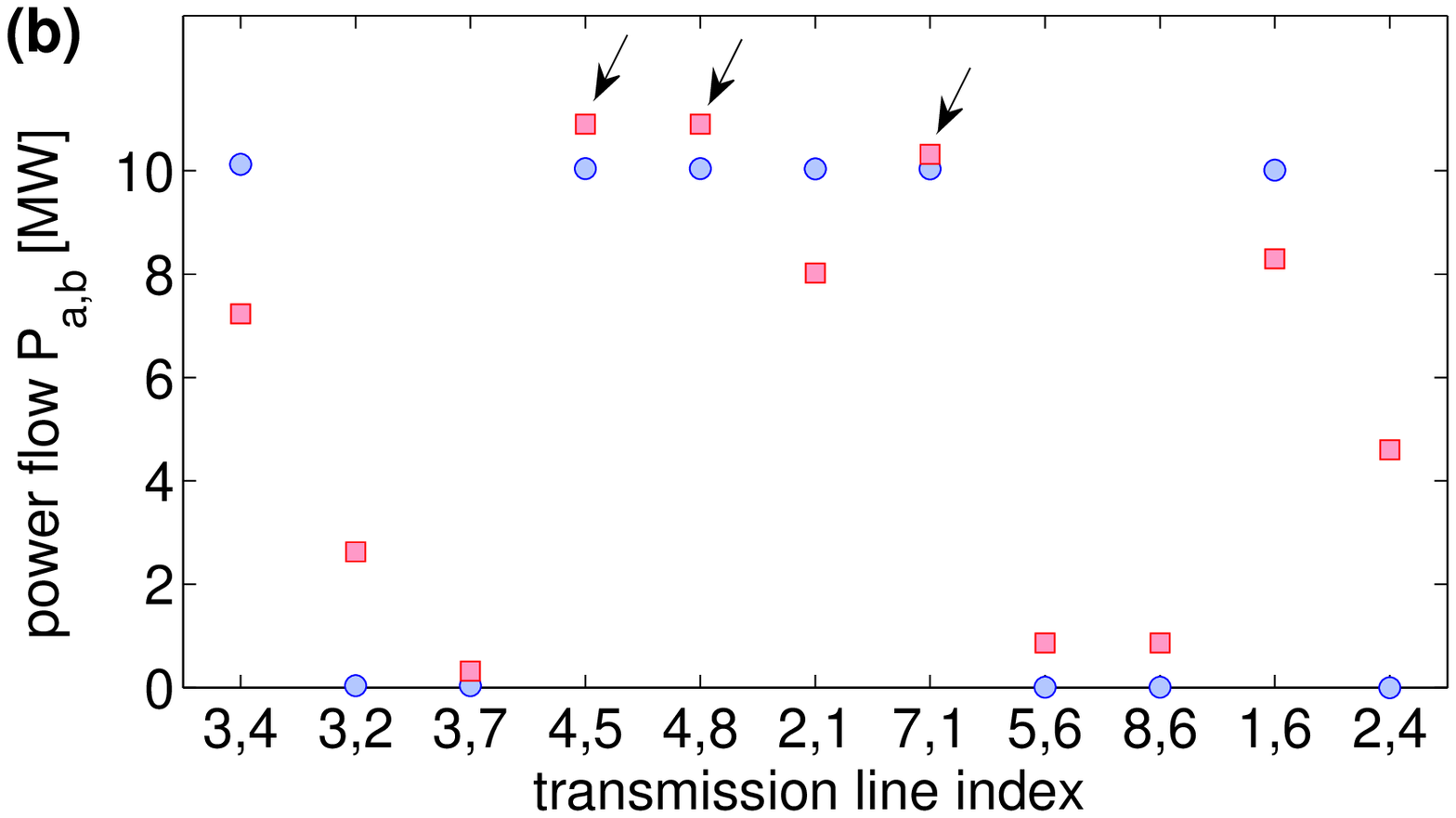}
\caption{\label{fig:pflow1}
(color online) Braess' paradox in an AC power grid in static operation.
(\textbf{a}) Network structure of the power grid including
four consumers, three generators and a slack node. The
detailed parameters are given in the text.
(\textbf{b}) Power flow along each transmission line without
(blue circles) and with the additional line (red squares). 
The transmission lines marked by an arrow are subject to
Braess' paradox: The maximum load $P_{4,5} = P_{4,8}$ 
\dirk{and $P_{7,1}$ increase} 
when the new line is added to the power grid.
The actual load of several transmission lines is given
in Tab.~\ref{tab:pflow1}.
}
\end{figure} 

As above we consider a power grid consisting of four 
generators and four identical consumers, 
cf.~Fig.~\ref{fig:pflow1} (a), assuming a 
nominal grid voltage of $110 \, {\rm kV}$.
One generator is chosen as slack node, while
the remaining ones provide the fixed real power $P_{\rm gen}$
at the grid voltage such that $|U_{\rm gen}| = 110 \, {\rm kV}$.
Each consumer node consumes a fixed real power
$P_{\rm con} = -10 \, {\rm MW}$ and a reactive 
power $Q_{\rm con} = -3.3 \, {\rm MVar}$. Hence, 
the power factor of the consumers is 
$\lambda = P_{\rm con}/\sqrt{P_{\rm con}^2 + Q_{\rm con}^2} \approx 0.95$.
We assume that the transmission lines are inductive and
suffer from ohmic losses, setting 
$Z = (10 + 20i) \,  \Omega$.

When the transmission line $2 \leftrightarrow 4$ is put into
operation, this can cause Braess' paradox for the connecting
lines. Figure \ref{fig:pflow1} and Tab.~\ref{tab:pflow1} show
the power loads in the network without (blue circle) and with 
(red squares) the new transmission line. The load of the 
connecting lines $4 \leftrightarrow 5$ and 
$4 \leftrightarrow 8$ \emph{increases} by $8.5$ \%.

We remark that the effective loss in the network, i.e. the
difference of the power generated and consumed at the nodes,
will never increase when the new transmission line is put into 
operation. From this point of view, building new lines is always 
favorable. Still, the maximum power load of single links in 
the network can be \emph{increased}, decreasing the margin 
to outage. In extreme situations, this may cause a 
shutdown of an overloaded transmission line and finally
a cascade of failures leading to a major power outage.

\begin{table}
\begin{tabular}{c | c | c | }
& initial grid  & with additional line \\
\hline
$P_{\rm slack}$               &    $10.191$ MW   &   $10.175$ MW       \\
$P_{4,5} = P_{4,8}$      &    $10.04$ MW   &   $10.90$ MW  \\
$P_{3,4}$                        &    $10.12$ MW   &   $7.23$ MW    \\
$P_{2,4}$                        &    $0$ MW   &   $4.60$ MW    \\
\hline
\end{tabular}
\caption{
\label{tab:pflow1}
Power generation and load of selected transmission lines 
for the power grid shown in Fig.~\ref{fig:pflow1}.
}
\end{table}

\subsection{Braess' paradox and desynchronization in oscillator networks}

As demonstrated recently \cite{12braess}, Braess' paradox also exists in oscillator networks. In particular, networks of two-variable oscillators describing 
the \emph{dynamics} of AC power grids and thus going beyond the static regime analyzed so far, typically exhibit Braess paradox for at least a fraction of (potentially) added links. The class of oscillator models  \cite{Fila08,12powergrid} is particularly appealing as it captures several collective phenomena present in real power grids whereas it is simple enough to admit a mechanistic understanding of such phenomena and simulations also for large complex networks.

\begin{figure}[t]
\centering
\includegraphics[width=7.5cm, angle=0]{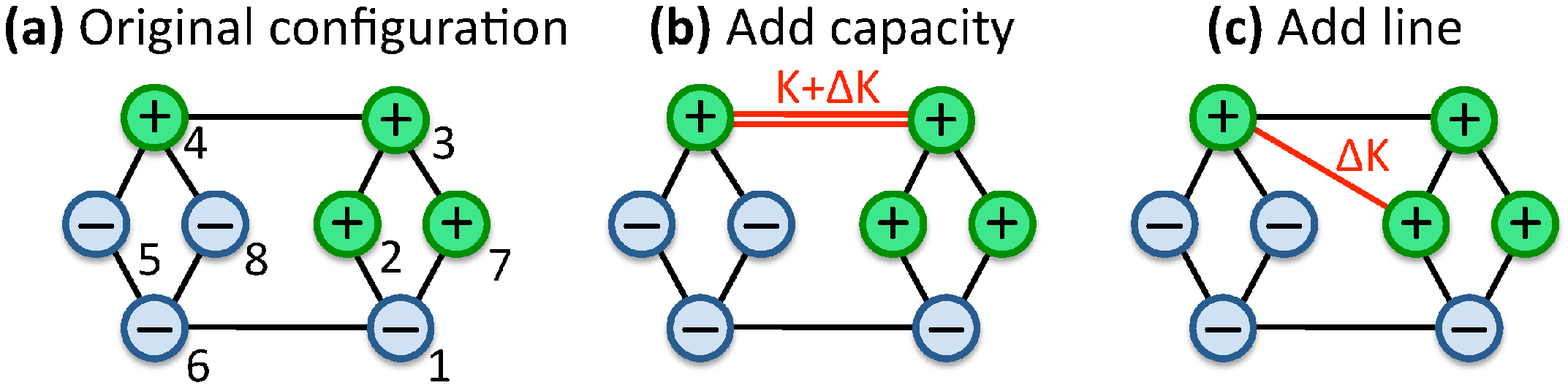}
\includegraphics[width=8cm, angle=0]{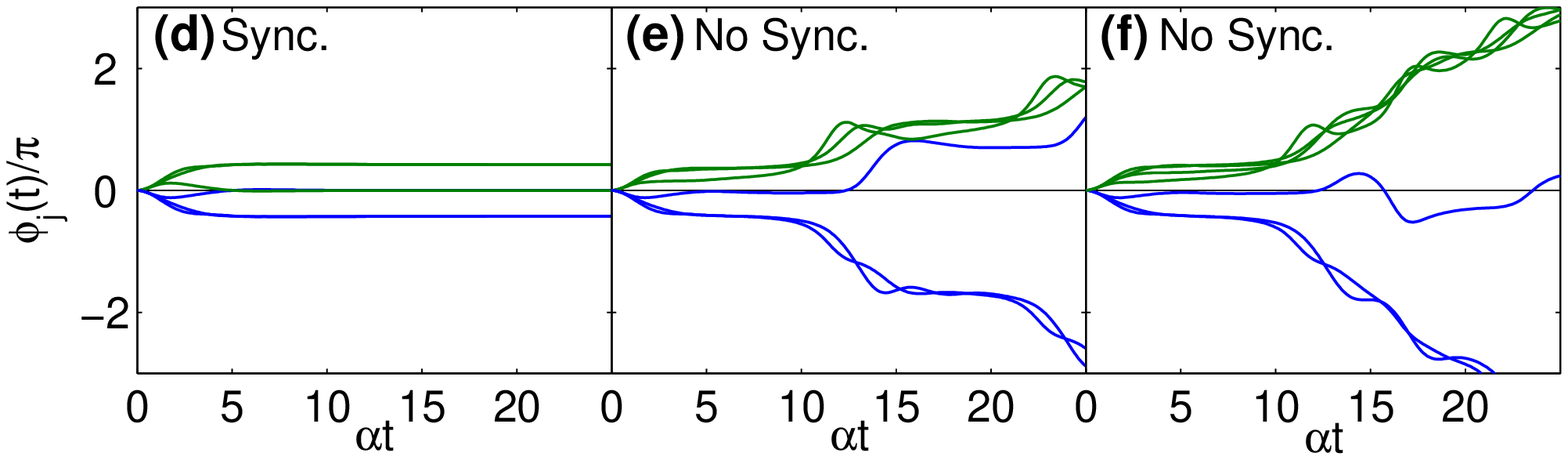}
\caption{\label{fig:net1}
(color online) Braess' paradox in an oscillator model \cite{12braess}
(a-c), Topology of the network. The vertices generate/consume
the power $P_j = \pm P$. The transmission lines have a capacity $K$.
(d) The original network converges to a phase-locked state.
When the capacity of one link is doubled (e), or when
a new link is added to the network (f), the steady state
ceases to exist and phase synchronization breaks down.
Parameters are $K = 1.03 \, P$, 
\dirk{$\Delta K = K$,}
$\alpha = P$, and the initial 
conditions are  $\phi_j = \dot \phi_j = 0$. 
}
\end{figure} 

 \begin{figure*}[tb]
\centering
 \begin{minipage}[b]{8cm}
 	\centering
 	\includegraphics[width=\textwidth]{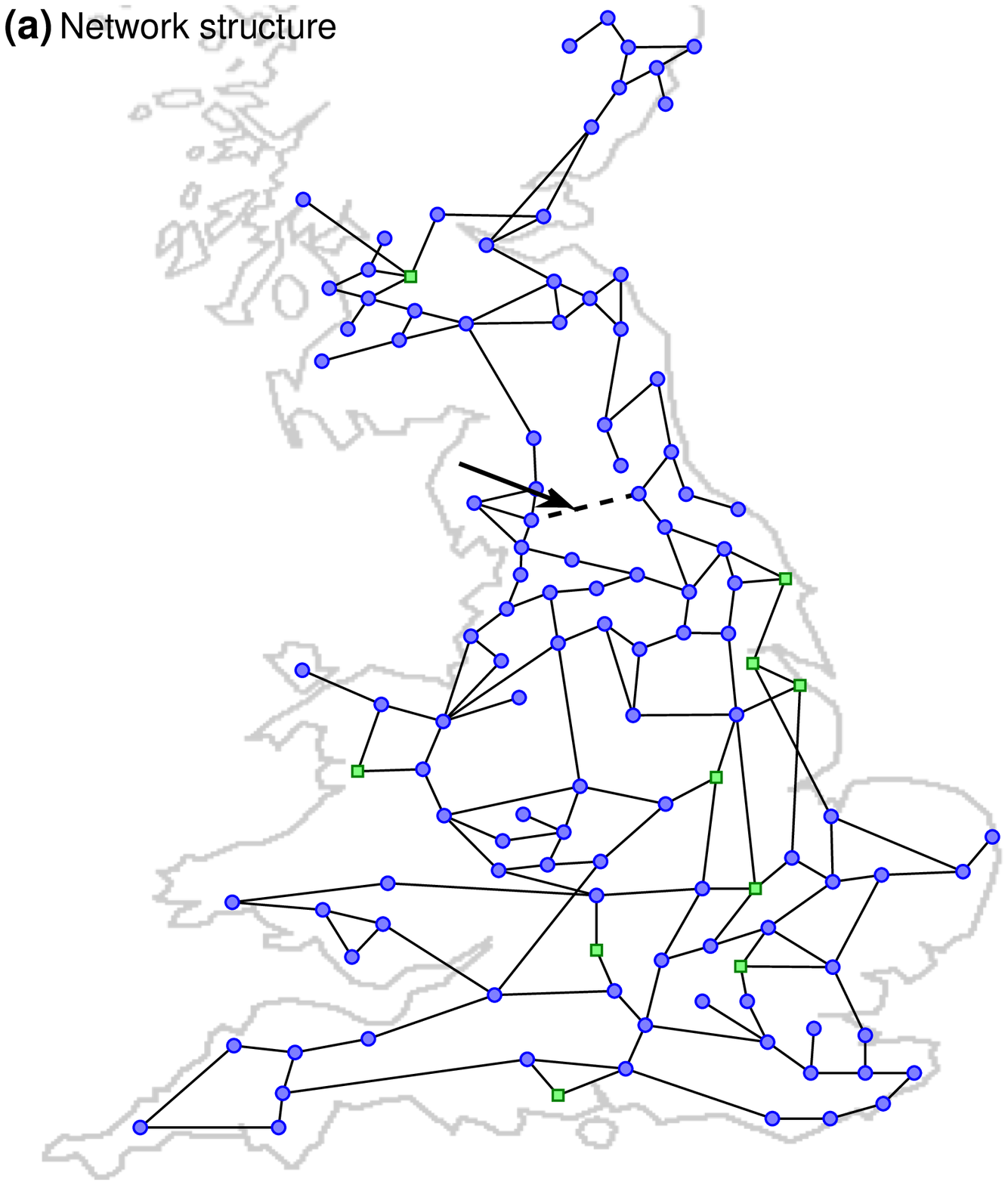}
\end{minipage}
\hspace{2mm}
\begin{minipage}[b]{7.7cm}
          \centering
	\includegraphics[width=\textwidth]{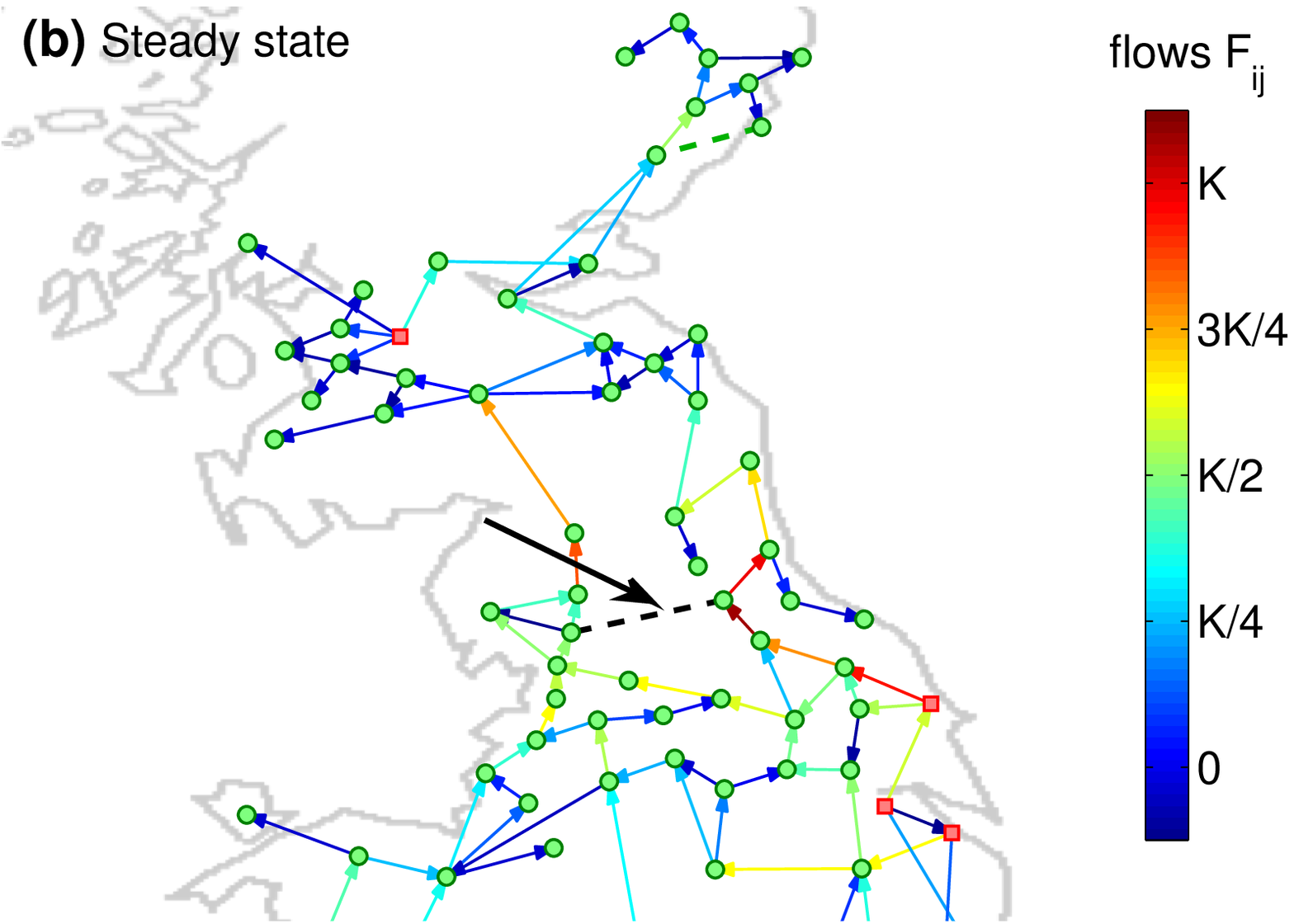}
	\includegraphics[width=\textwidth]{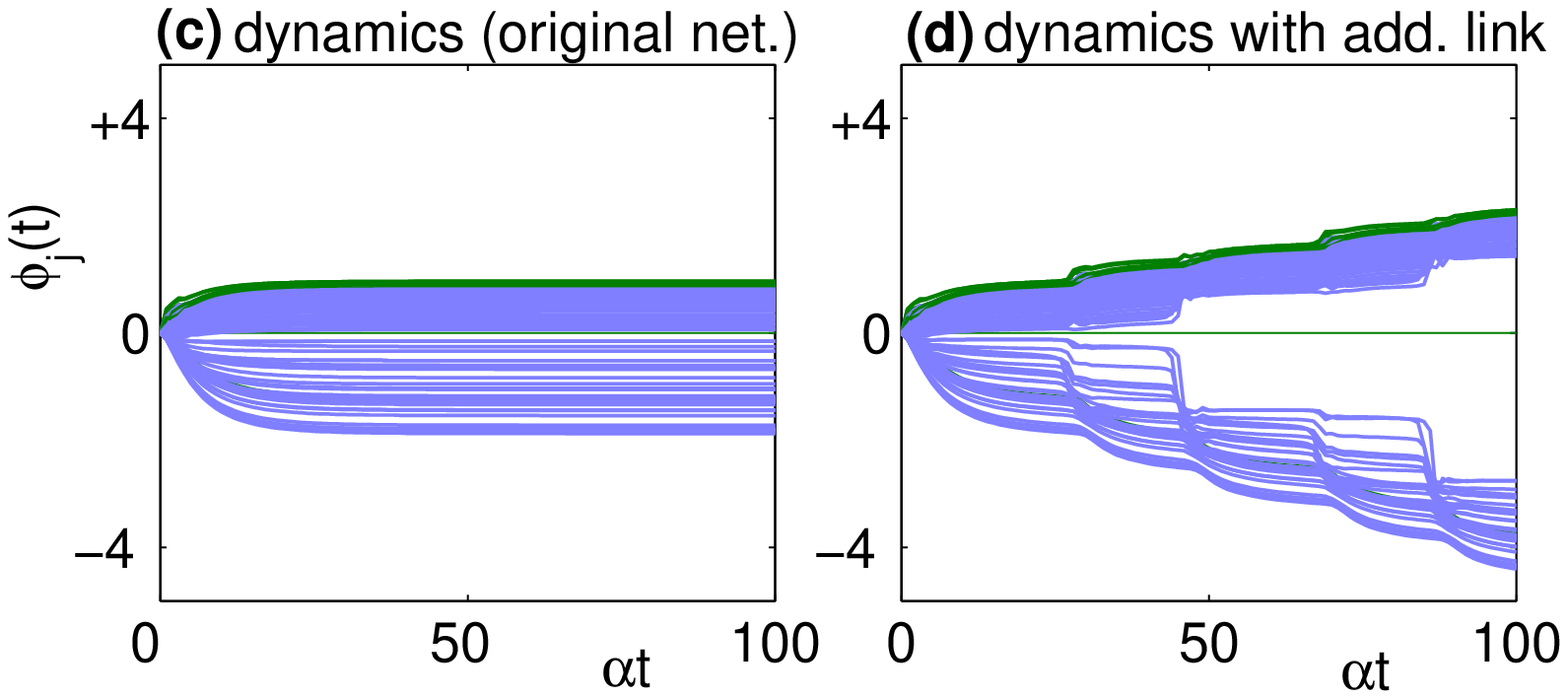}
\end{minipage}
\caption{\label{fig:uksync}
Desynchronization due to Braess' paradox in an oscillatory power grid with complex topology \cite{12braess}.
(\textbf{a}) Topology of the British power grid, consisting of 120 nodes and 
165 transmission lines (thin black lines) \cite{Simo08}. 
Ten nodes are randomly selected to be generators 
($P_j = 11 P_0$, $\square$), the others are consumers 
($P_j = -P_0$, $\circ$). The dashed new link causes 
desynchronization due to Braess' paradox.
(\textbf{b}) Power flow in the original network for $K=13 P_0$.
The load of several transmission lines in the neighborhood of the new 
link is already close to the maximum capacity $K$.
(\textbf{c}) For $K = 13 \, P_0$ and $\alpha = P_0$, the initial network 
converges to a phase-locked steady state.
(\textbf{d}) Synchronization becomes impossible
after the new link has been added.
The initial condition is fully synchrony, $\phi_j = \dot \phi_j = 0$.
}
\end{figure*}

In this model, both generators and consumers are assumed to be synchronous 
machines and thus obey the same equation of motion with a parameter $P$ giving 
the generated $(P>0)$ or consumed $(P<0)$ power. 
The state of each machine is determined 
by its phase angle $\theta(t)$ and velocity $\dot\theta(t)$. The mechanical phase of
the $j$th machine is written as $\theta_j(t) = \omega_0 t + \phi_j(t)$, where
$\omega_0 = 2 \pi \times 50 \, {\rm s}^{-1}$ or
$\omega_0 = 2 \pi \times 60 \, {\rm s}^{-1}$
is the reference frequency of
the power grid. The equation of motion is obtained via the principle of energy
conservation, that is the generated or consumed power of each element must 
equal the power exchanged with the grid plus the accumulated and the dissipated 
power. The power transmitted between machines $i$ and $j$ is proportional to
the sine of the phase difference and the capacity of the transmission line $K_{ij}$.
As shown in detail in appendix \ref{sup:oscillator model},
the equations of motion for the phase differences then read
\begin{equation}
   \frac{d^2 \phi_j}{dt^2} = P_j - \alpha \frac{d \phi_j}{dt}
          + \sum \nolimits_i K_{ij}\sin(\phi_i-\phi_j).
        \label{eqn:eom}  
\end{equation}

Again we first consider an analytically solvable model network which is similar
to the ones discussed in the previous sections (cf. Fig.~\ref{fig:net1}). Four 
generators with $P_j = + P$ and four consumers with $P_j = -P$ are connected 
by transmission lines with capacity $K$. A stable steady state 
exists for the original network structure (Fig.~\ref{fig:net1} (a)), 
\dirk{if the power is smaller than a critical value
\be
  P \le P_{\rm max} = K,
  \label{eqn_Kcrit}
\ee
} and the system rapidly relaxes to this phase-locked state of partial 
synchrony, cf.~Fig.~\ref{fig:net1} (d).
As in the previous examples, the increase of the transmission capacity
or the addition of a new link can induce Braess' paradox. In these cases
synchronization becomes impossible -- the phases $\phi_j(t)$ cannot
phase lock as shown in Fig.~\ref{fig:net1} (e,f).
For a real power grid this effect would imply the automatic shutdown 
of the desynchronized generators, which may then cause a major
power outage in the complete network.

The loss of synchrony in an oscillator network due to Braess' paradox
is a rather subtle effect. The condition for the existence of a phase-locked 
steady state in the power grid model $\phi_j = \dot \phi_j = 0$ is equivalent
to the conservation of the flow
\be
   \sum_{j=1}^{N} F_{ij}  = P_i \qquad \forall i = 1, \ldots , N_{n}.
   \label{eqn:flowcon-osc}
\ee
For the oscillator model, the power flows are given by
\be
   F_{ij} = K_{ij} \sin(\phi_j - \phi_i).
\ee
\dirk{Therefore, we can divide the condition for the existence of a 
phase-locked steady state into a dynamic and a geometric component.}
In addition to the conservation of the flow (\ref{eqn:flowcon-osc}) we  have 
to satisfy \dirk{a geometric} condition: For every cyclic path in the network, 
the sum of all phase differences  must vanish such that all phases are well-defined,
\be
    \sum \nolimits' (\phi_j - \phi_{i}) =  
    \sum \nolimits'  \arcsin(F_{ji}/K_{ji}) = 0 \quad (\text{mod} \, 2 \pi).
    \label{eqn:phase_cycle}
\ee
The prime indicates that the sum is taken along a cyclic path. 
For the networks shown in Fig.~\ref{fig:net1} one can easily find values
$F_{ij}$ which satisfy flow conservation, cf.~Eq.~(\ref{eqn:flows-ss}),
but the condition (\ref{eqn:phase_cycle}) is no longer fulfilled.
For such $F_{ij}$'s, despite the fact that all dynamical conditions (\ref{eqn:flowcon-osc}) 
are satisfied, no steady state exists due to \emph{geometric frustration}, the incapability of the system to satisfy (\ref{eqn:phase_cycle}) along all cycles of the network.
In fact, the critical coupling strength $K_c$ for the existence of a 
phase-locked state is increased. Thus geometric frustration limits the 
capability of the network to support a steady state. 

Braess' paradox in oscillator networks is rooted in the geometric 
frustration of small cycles, which are generally present in most 
complex networks \cite{12braess}. We speculate that condition (\ref{eqn:phase_cycle}) is more often satisfied along long cycles because these have a larger number of variables, i.e. the restriction can be ``solved'' in higher-dimensional space of phase differences. Braess' paradox occurs in many, but not all complex networks as elementary cycles are typically overlapping such that the effects of geometric frustration depend on the precise network topology and are as such hard to predict.

\begin{figure*}[tb]
\centering
\includegraphics[width=16cm, angle=0]{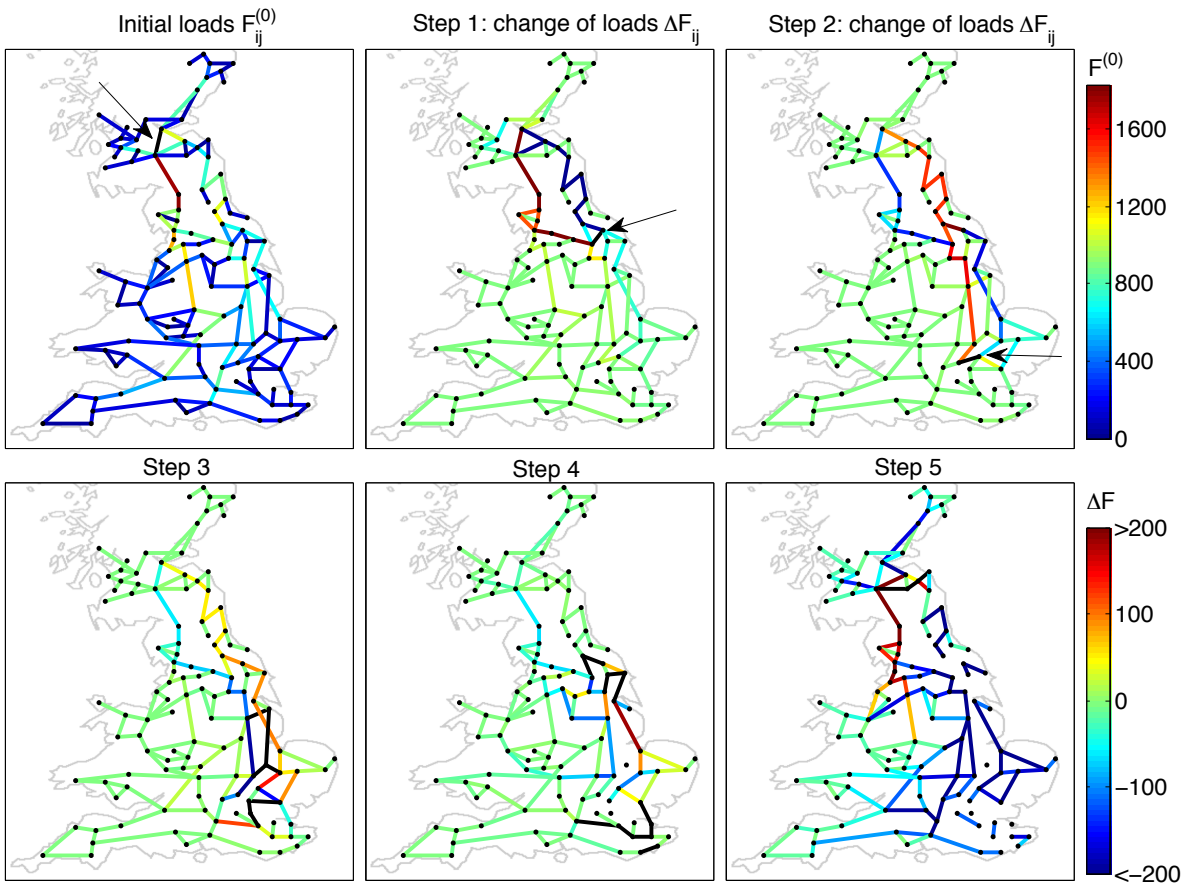}
\caption{\label{fig:cascade1}
A cascading failure triggered by the addition of a new link in the messaging model.
Upper left panel: The initial network and the loads \dirk{$F_{ij}^{(0)}$} of all links 
in a color code. The link colored in black is added to the network.
The remaining panels show how the failure propagates through the network.
The color code gives the change of the load \dirk{$\Delta F$} in comparison 
to the previous
step. Links colored in black are overloaded and drop out of service.
Note that in this example links are overloaded which are \emph{not} nearest
neighbors of the additional link which triggers the cascading failure. These links 
are marked by arrows in step 1 and 2.
In this simulation we have used the coarse grained structure of the British
power grid with 120 nodes and 165 edges as described in \cite{Simo08}.
The tolerance parameter is $\alpha = 0.32$. 
}
\end{figure*}

As an important example, we consider the British high-voltage power 
transmission grid shown in Fig.~\ref{fig:uksync}, cf.~\cite{Simo08}. In our 
study, we randomly choose ten out of 120 nodes to be generators 
($P_j = + 11 P_0$), while the remaining ones are consumers
($P_j = - P_0$). For $K=13 P_0$ the original network relaxes to a phase-locked 
steady state as shown in part (c) of the figure. If one inappropriate new link 
is added (dashed black line), global synchronization (phase locking) is lost due 
to Braess' paradox  for the given coupling strength $K$. Instead, the power 
grid decomposes into two asynchronous fragments as shown in part (d) of the figure.

Furthermore, this example emphasizes the importance of the lines connecting to the newly added one: These lines may easily become overloaded when the new edge is added 
to the network, finally causing a system wide failure. In particular, also lines indirectly connecting to the newly added one may be affected in a similar way. Figure~\ref{fig:uksync} \dirk{(b)} shows the power flow in the 
steady state before the new link is added. One observes that several transmission
lines in the neighborhood of the new link are heavily loaded. These lines get
overloaded when the new link is put into operation, causing the 
desynchronization of the grid. 
Generally, adding new transmission lines are assumed to most strongly change the load distribution in the network if they 
are build in regions where the existing lines are already heavily loaded. In fact, loads are often thought to be reduced due to the new line. However, load redistribution is not necessarily supportive, as shown above.
For more details on the physics of Braess' paradox in oscillator networks, see also \cite{12braess}.

\section{Cascading failures triggered by the addition of links}
\label{sec:cascade}

The messaging model studied in Sec.~\ref{sec:braess-motlai} was initially
introduced to study how the damage of a \emph{single} link can induce 
a major failure in large parts of the grid \cite{Mott02,Simo08}.  
This scenario is very important for real supply networks, in particular
for power grids, as most major power outages are the result of a cascade
of failures. A prominent example is the power outage in the western 
European power grid on November 4th 2006, which was triggered 
by the shutdown of one double-circuit transmission lines over the river Ems in 
north-western Germany \cite{UCTE07}. As a consequence ten million households 
were disconnected from the power supply. Power outages even 
occurred in Spain, approximately 2000 km away from the cause
of the cascading failure.
In this section we show that not only the removal or damage of at single
transmission line, but also the \emph{addition} of a new transmission line
can cause a cascading failure due to Braess' paradox.

An example for a cascading failure by link addition is shown in Fig.~\ref{fig:cascade1}.
Again, we consider the coarse grained structure of the British power grid with
120 nodes and 165 link \cite{Simo08}. The upper left panel shows the load
$F^{(0)}_{ij}$ of each link of the initial network in a colormap plot. As in the 
original model \cite{Mott02}, we assume that the capacity of 
each link is adapted 
to the loads
\be
   K_{ij} = (1+\alpha) F^{(0)}_{ij} 
\ee
with a tolerance parameter $\alpha \ge 0$. If the load $F_{ij}$ exceeds the
capacity $K_{ij}$, then the link becomes overloaded and drops out of service,
i.e. it is removed from the network.

Now we consider the effect of the \emph{addition} of a new link to the network.
The new link is colored in black and marked by an arrow in in the upper left 
panel of Fig.~\ref{fig:cascade1}. We assume that the new link has a rather
high capacity $K_{\rm new} = \max_{ij}(K_{ij})$.
The upper middle panel then shows how the loads of all links changes after 
the addition of this link. As expected from the study of the elementary model
in Sec.~\ref{sec:braess-motlai}, the load of the \emph{connecting lines} 
along the western coast increases. For $\alpha = 0.32$, only a single link
becomes overloaded which is colored in black and marked by an arrow.
This failure then again causes a redistribution of the loads in the network
and the overload of another link. This finally triggers a whole cascade of failure
as shown in the remaining panels of Fig.~\ref{fig:cascade1}.
After six steps, the cascade stops and the network is decomposed 
into 13 different components. The largest
component includes 85 nodes, while 6 nodes are completely disconnected.

\begin{figure}[t]
\centering
\includegraphics[width=8cm, angle=0]{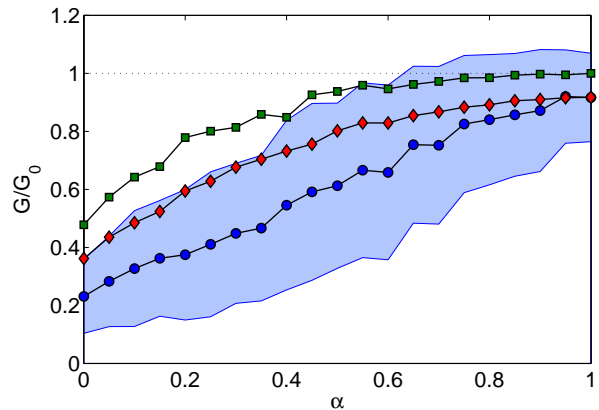}
\caption{\label{fig:cascade2}
(color online) How harmful are single link changes? Characterization of the final state after a cascading failure 
in the messaging model. Plotted is the relative size of the 
largest connected component $G/G_0$ in the final state,
i.e. the final size divided by the initial size as a function of 
the tolerance parameter $\alpha$.
We consider the addition of links at random positions 
(blue circles), the addition of links at local position 
(green squares) and the removal of links (red diamonds).
Local position means that a link is added between two 
random nodes with initial graph distance of 2.
In this simulation we have used the coarse grained 
structure of the British power grid with 120 nodes and 
165 edges as described in \cite{Simo08}.
Results have been averaged over 200 random positions 
(for local and global random addition of links) and over all 
165 existing links (removal of links). The shaded (blue) area 
shows the standard deviation for the cases of link addition
at random position. 
}
\end{figure}

An important question is how crucial the addition of a new link is in general,
in particular in comparison to the removal of a single links. To answer this
question, we have simulated the impact of cascading failures as a function of
the tolerance parameter $\alpha$ for the coarse grained structure of the British
power grid. A single link is added or removed at random positions in the network
and we analyze the final structure of the network after the cascade of failures
has come to an end.
Figure \ref{fig:cascade2} shows the size of the largest connected component $G$ 
relative to the size of the initial network. 
The simulations reveal a surprising result: Link addition has even more severe 
consequences than link removal in the sense that it leads to smaller values of 
$G/G_0$. However, this effect is partly due the fact that links are added at 
completely random positions such that they generally connect \emph{distant} areas
of the network. Obviously, this can have a stronger effect on the network flow
than the 'local' removal of a link.
The addition of links at random positions is also rather unrealistic. Thus we also
consider an alternative scenario where links can be added only between nodes
with a distance of two. The numerical results (Fig.~\ref{fig:cascade2},
green squares) reveal that such a local addition of links has a less severe effect
than the removal of links, but both can be substantial.

\section{Nonlocal impact of link addition and removal}
\label{sec:nonlocal}

Cascading failure events such as those illustrated in Fig.~\ref{fig:cascade1} 
highlight that a single link addition may induce strong
nonlocal impacts. In the example (Fig.~\ref{fig:cascade1}), the cascade is triggered by the addition of a link in Scotland.
The load of the connecting links is strongly modified, also beyond the 
immediate neighborhood of the novel link. In particular, a rather weak 
link located in northeastern England becomes overloaded, while the 
links in the immediate neighborhood remain in operation. In the second
step, one link at an even further distance becomes overloaded.
Finally, several nodes in southeastern England are fully disconnected 
from the network -- nodes which are far away from the link which
caused the failure.

\begin{figure}[t]
\centering
\includegraphics[width=8cm, angle=0]{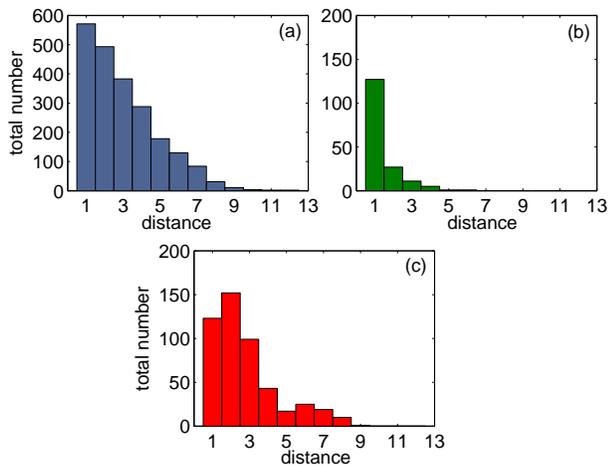}
\caption{\label{fig:nonloc1}
Nonlocal impact of link addition and removal (here: in the messaging model).
(a-c) Histograms of the distances of the initially added/removed link
to the links which become overloaded the first step.
We consider (a) the addition of links at random positions, (b) 
the addition of links at local position and (c) the removal of links.
Note that the distance is defined in terms of the original unmodified network 
in the case of link removal. For link addition, distance is defined in
terms of the modified network, i.e. the network including the additional
link.
In this simulation we have used the coarse grained structure of the British
power grid with 120 nodes and 165 edges as described in \cite{Simo08}.
Results have been collected for 200 random positions.
}
\end{figure} 

The example shown in Fig.~\ref{fig:cascade1} is surely extreme, but a strong 
nonlocal impact triggered by changes of the network topology is by no means 
exceptional:
To analyze the geographic properties of cascading failures quantitatively,
we calculate the distance between the overloaded links and the added/removed 
link, which causes the overload. 
Figure \ref{fig:nonloc1} shows a histogram of these distance for the first step
of the cascade. As above, we consider three scenarios: (a) the addition of a link
at a random position, (b) the local addition of a link and (c) the removal of a link.
For scenarios (a) and (c) it is observed that overloads frequently occur at remote 
positions. In the case of link removal, the next-to-nearest neighbors are even 
more vulnerable to an overload than the nearest neighbors.
If link addition is restricted to a local position (scenario b), then also the immediate 
impact occurs predominantly at local positions. However, the dynamic
consequences often reach beyond the local neighborhood in the network.
Already in the second step of the cascade (not shown), almost no differences
can be observed between the three different scenarios.

We note that these findings do not contradict the claim that the connecting lines
are crucial for the possibility of Braess' paradox. They rather show that we have to
consider the connecting lines beyond the immediate neighborhood, too.
In particular, in the example studied in Fig.~\ref{fig:cascade1}, one can identify a rather
long path of connecting lines, whose load strongly increases after the addition 
of the new link (the red link in the upper middle panel in Fig.~\ref{fig:cascade1}).
The overload then occurs on the weakest of these links, not on the nearest.

\section{Braess' paradox on complex network topologies}
\label{sec:complex}

\begin{figure}[t]
\centering
\includegraphics[width=8cm, angle=0]{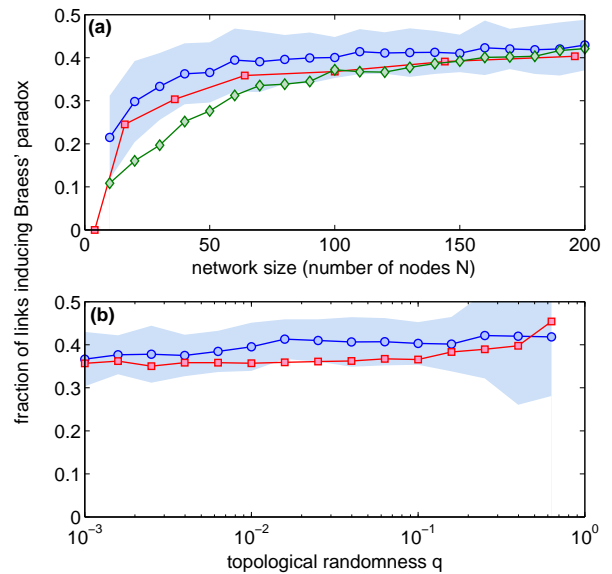}
\caption{\label{fig:bflow_nq}
(color online) Prevalence of Braess' paradox across different topologies:
Randomized ring lattices (blue circles, \cite{Watt98}), randomized 
square lattices (red squares, \cite{Klei00}) and networks with
algebraic degree distribution (green diamonds, \cite{Bara99}).
Plotted is the fraction of links whose removal leads to a \emph{decrease}
of the maximum flow $F_{\rm max}$ as a function of (a) the network 
size 
\dirk{($q=0.1$)} 
and (b) the topological randomness of the network ($N=200$).
The shaded blue area shows the standard deviation for the randomized
ring lattices. Solid lines are drawn to guide the eye.
Random networks which are not globally connected were discarded, in particular there is no data point at $q=1$ as random networks with average degree $d=4$ are almost always disconnected.
}
\end{figure} 

Braess' paradox arises for different models of supply networks and 
can have severe consequences for the operation of the network. We 
now show that in large complex networks this phenomenon is the rule 
rather than the exception. 
We systematically study the occurrence of Braess' paradox for the flow
model introduced in Sec.~\ref{sec:braess_flow} and analyze how this 
behavior depends on \dirk{the size and topology} of the network.

\dirk{To analyze how topological changes impact network dynamics for
different classes of random networks, we first consider} 
classes of networks that interpolate between regular and 
random topology, also referred to as small-world networks
\cite{Watt98,Klei00}. Starting with a ring lattice where each node
is connected to its nearest and next-to-nearest neighbors or a 
square lattice, every link is rewired with probability $q$, i.e. removed 
and re-inserted at a different randomly chosen position. 
Furthermore we consider networks with an algebraic degree distribution
generated by random preferential attachment \cite{Bara99}.
The average degree is $d=4$ for the randomized lattices and 
$d = 4 - 4/N \approx 4$ for the algebraic network.
In each case, half of the nodes are randomly chosen to be generators 
($P_j= +P_0$) or consumers ($P_j= - P_0$).

We then check for each link how its removal affects the maximum flow 
$F_{\rm max} = \max_{ij} |F_{ij}|$ in the network. A link is said to induce
Braess' paradox if $F_{\rm max}$ \emph{decreases} after the removal.
This definition is exactly the same as above as the re-addition of the link
would increase $F_{\rm max}$ which could then cause a fatal overload.
For each random network we count the number of links which 
induce Braess' paradox and average over one-hundred realizations 
of the network structure. We then analyze how this number depends
on the size and topology of the networks.
  
\begin{figure}[t]
\centering
\includegraphics[width=8cm, angle=0]{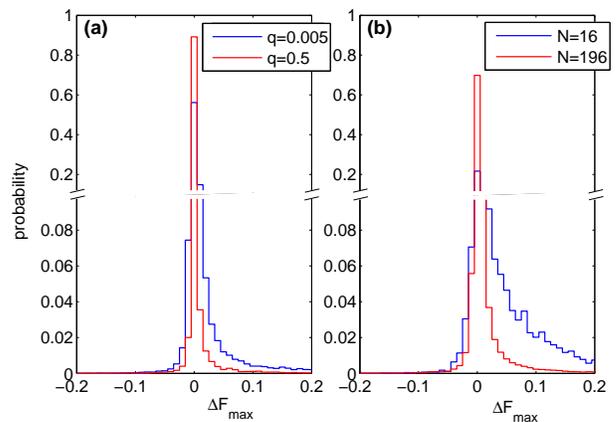}
\caption{\label{fig:bflow_hist}
(color online) Impact of link removal on the maximum flow in a supply network.
Histogram displays the difference of the maximum flow 
\dirk{$\Delta F_{\rm max}$} before
and after link removal  (\ref{eqn:dflow}) for randomized square
lattices \cite{Klei00} with
(a) $N=100$ and two different values of $q$ and
(b) $q=0.1$ and two different values of $N$.
A negative difference  $\Delta F_{\rm max}$ indicates Braess' paradox.
}
\end{figure}

The numerical results plotted in Fig.~\ref{fig:bflow_nq} demonstrate that 
Braess' paradox is common in complex flow networks.
For large networks with 200 nodes, more than 40 \% of the links 
impede the operation of the the supply network
in the way that their removal is beneficial as it decreases the maximum 
load $F_{\rm max}$.
However, the impact of a single link decreases with the total number
of links present in the network. Therefore, the magnitude of the change
of $F_{\rm max}$ is generally smaller for larger networks, see Fig.~\ref{fig:bflow_hist} (b) where we have plotted a histogram for the difference of the maximum
loads before and after the link removal
\be
   \Delta F_{\rm max} = F_{\rm max, after}  -  F_{\rm max, before}.
   \label{eqn:dflow}
\ee  
for a small ($N=16$) and a large ($N=196$) network. One clearly sees that 
the probability for large changes $\Delta F_{\rm max}$ is extremely small
for the larger network. 

Furthermore, Braess' paradox prevails across all types of
network topologies and model dynamics considered here. For large networks, the topology has 
only a minor influence on the probability that a link induces Braess'
paradox. Even more, the probability is almost independent of the 
topological randomness $q$ as shown in Fig.~\ref{fig:bflow_nq} (b).
However, topological changes have a more severe impact in networks
with a regular topology as shown in Fig.~\ref{fig:bflow_hist} (a), where
we compare the difference of the maximum flow $\Delta F_{\rm max}$ before
and after links removal for a randomized square lattice with $q=0.005$
(close to regular) and $q=0.5$ (close to random). The probability for large values of $F_{\rm max}$ is much larger 
for the regular network with $q=0.005$.

\dirk{
In conclusion we find that the probability to observe Braess' paradox
is almost independent of the of the network topology and increases 
with the network size. However, the impact of a single link is largest
for regular structures and decreases with the network size.
This finding is consistent with the results for oscillator 
networks discussed in \cite{12braess}.
}

\section{Discussion}

We have revealed and analyzed Braess' paradox for different models
of supply networks: Against the naive intuition, the addition of new 
connections in a network does \emph{not} always increase the overall
transmission capacity of the network.
An elementary model system was designed to analytically study this 
paradoxical behavior, while numerical simulations have revealed that a 
substantial fraction of potential links induce this deleterious transition.
Furthermore, we have shown that catastrophic failures of complex supply 
networks may not only be caused by failures of 
single elements or links but also by the addition of  \emph{single} links. 
In particular, networks may destabilize due to a non-local overload.

Future model studies must include the detailed structure of supply
networks as well as the spatial and  temporal heterogeneity of generation
and demand. For instance, both the consumption and generation  of
electric energy in modern power grids is strongly fluctuating. 
This notwithstanding, Braess' paradox is a general feature such
that it can play a crucial  role in real supply networks.
In the future, it will be of great scientific as well as 
economic interest to understand how these phenomena depend 
on the topologies of the underlying networks in detail, 
cf.~\cite{Stro01,Barr04,Schn11,Vesp11}. In particular, a badly 
designed electric power grid subject to Braess' paradox may cause 
enormous costs for new but counterproductive electric power lines
that actually reduce grid performance and stability. 

While Braess' paradox has been observed in real-world traffic networks 
in several locations \cite{Cohe91,Roug02,Blum07,Youn08,Vali10},
its effects are less apparent for different types of supply networks. 
For instance, power grids are usually operated far from their load limit 
such that local structural changes do not cause obvious global failures.
However, periods of extreme load do occur and are expected to 
become more likely in future power grids with many strongly 
fluctuating renewable energy sources.
In such a period of extreme load, small local changes of the network
structure may cause a global breakdown. For instance, there was
a significant East-West power flows in the european power grid
on 4 November 2006 because of a large wind feed-in in Germany.
In this situtation, the disconnection of one double-circuit 
transmission line in Nothern Germany was sufficient to trigger 
a global power outage in most of western europe \cite{UCTE07}
A key result of this work is that in such a situation the \emph{addition} 
of a new link can be just as fatal as the disconnection of an existing one.

\acknowledgments

We thank S.~Grosskinsky, M.~Rohden, A.~Sorge, 
D.~Heide, and R.~Sollacher for valuable discussions. 
Supported by the Federal Ministry of Education and Research (BMBF)
Germany under grant number 01GQ1005B and by a grant of the
Max Planck Society to M.T.

\appendix

\section{Models of supply networks}
\label{sec:appendix_models}

In this appendix we provide a detailed description of the different models
of supply networks used in the present paper.

\subsection{Messaging model}
\label{sec:messagemodel}

A popular model to study the stability of supply and communication networks,
in particular the vulnerability to cascading failures, has been introduced by
Motter and Lai \cite{Mott02}. In contrast to the original study we focus
on the links in the network, not the nodes. In particular, we assume that a
link drops out of service if it is overloaded, while the nodes are not affected.

The messaging model assumes that at each time step, one unit of information or
energy is send from each node to each other node in the connected component 
along the shortest path.
The load of each link $F_{ij}$ is then given by the number of shortest paths 
running over this link $i \leftrightarrow j$, which is nothing than the edge
betweenness centrality. Furthermore,
it is assumed, that the capacity of each link proportional to the load of
the link in the initial intact network,
\be
   K_{ij} = (1+\alpha) F^{(0)}_{ij} \, .
\ee
Here, the superscript $(0)$ denotes the intact network.
Then Motter and Lai analyze what happens if one link is damaged. Obviously
the other links have to take over the load such that $F_{ij}$ will generally 
increase. If the load exceeds the capacity of a link, $F_{ij} > K_{ij}$, then this
link will also drop out of service, which can trigger a cascade of failures 
disconnecting the entire grid.

\subsection{Flow model}
\label{sec:flowmodel}

One of the simplest models of supply networks considers 
only the flow between different elements of the network. A similar 
model has also been used to model biological flow models in \cite{Kati10}.
The power grid is specified by the transmission capacity $K_{ij} > 0$
between the nodes $i,j \in \{1,\ldots, N \}$, where $N$ denotes the number of nodes in 
the network. Obviously, we have $K_{ij} = K_{ji}$ and we set $K_{ij} = 0$ if no link
exists between nodes $i$ and $j$. Furthermore, each node of the network is characterized 
by the electric power $P_j$ it generates ($P_j > 0$) or consumes  ($P_j < 0$). 

We denote the flow from node $i$ to node $j$ by $F_{ij}$, 
which can be positive (power flows from $i$ to $j$) or negative 
(power flows from $j$ to $i$). 
The conservation of energy then directly leads to the condition
\be
   \sum_{j=1}^{N} F_{ij}  = P_i \qquad 
   \text{for all } j \in \{1, \ldots , N \}.
   \label{eqn:energycon2}
\ee
\dirk{In general, energy conservation is not sufficient to uniquely 
fix the flows, as it poses only $N-1$ linearly independent constraints
for the $L$ independent non-zero variables $F_{ij}$, $L$ being the 
number of links in the network.}

The unique steady state is determined by the condition that the total dissipated power
\be
    E_{\rm diss} =  \sum \nolimits^{'}_{i<j} \frac{F_{ij}^2}{2 K_{ij}}
\ee
should be minimal. In this expression, the primed sum runs only over existing
transmission lines, i.e. only over links with $K_{ij} \neq 0$. To minimize this
expression respecting the conservation of energy, we use the method of 
Lagrangian multipliers, i.e. we minimize 
\bea
   L &=& \sum \nolimits_{i<j} \frac{F_{ij}^2}{2 K_{ij}}
              - \sum_i \lambda_i  \left(  \sum_{j} F_{ij} - P_i  \right) \nn \\
      &=& \sum  \nolimits_{i<j} \left( \frac{F_{ij}^2}{2 K_{ij}} - (\lambda_i - \lambda_j) F_{ij} \right).
\eea
Minimization yields the condition
\bea
  && \frac{\partial L}{\partial F_{ij}}  = K_{ij}^{-1} F_{ij} - (\lambda_i - \lambda_j) \stackrel{!}{=} 0 \nn \\
  && \Rightarrow F_{ij} = K_{ij} (\lambda_i - \lambda_j)
\eea
Thus, the flow from $i$ to $j$ is given by a potential difference $\lambda_i - \lambda_j$
multiplied by the transmission capacity $K_{ij}$. 
Substituting this result  into equation (\ref{eqn:energycon2}), we find that the potential
is determined by a linear system of equations
\be
   \sum_j K_{ij} (\lambda_i - \lambda_j) = P_i .
\ee
for convenience, we rewrite this system of equations in a vectorial form
\be
   \widetilde K \vec \lambda  = - \vec P 
\ee
where we have defined the matrix elements
\be
   \widetilde K_{ij} = K_{ij} - \big( \sum \nolimits_k K_{ik} \big) \delta_{ij}
\ee
and the vectors $\vec P$ and $\vec \lambda$ collect the values of the power
generated/consumed in each node $P_j$ and the potentials $\lambda_j$.


We note that the oscillator model (\ref{eqn:osc-model}) reduces to this model
in the steady state, when the phase differences are so small
that one can approximate $\sin(\phi_i - \phi_j) \approx \phi_i - \phi_j$,
which is typically the case if the \dirk{couplings $K_{ij}$} are very large. The power
flow model can thus be seen as a limiting case of the oscillator model.

\subsection{Static flow in AC power grids}
\label{sup:ACflow}

In an AC power grid, one has to take into account that not only real 
but also reactive power is transmitted via the network. Each link of
the grid is characterized by its complex impedance, not just 
\dirk{its maximum transmission capacity.}
In a static power flow study one calculates the 
voltage of each node such that the electric power is conserved 
at each node \cite{Grai94}.

Every node $a$ is characterized by its voltage $U_a$ compared to the ground. 
A transmission line between the nodes $a$ and $b$ carries the current
\be
   I_{a,b} = \frac{1}{Z_{a,b}} (U_a - U_b),
  \label{eqn:acflow-current}
\ee
where $Z_{a,b}$ is the impedance of the line. The electric power transmitted 
from or two a node of the network is then given by
\be
   S_{a} = \sum_{b} S_{a,b} = \sum_b 3 U_a I_{a,b}^* ,
   \label{eqn:acflow-power}
\ee
where the star denotes complex conjugation. The real part gives the
real transmitted power $P_a$, while the imaginary part $Q_a$ is the 
reactive power,
\be
   S_a = P_a + i Q_a \, .
\ee
Power conservation requires that the total transmitted power equals 
the power generated or consumed at the respective node:
\be
    S_{a} = S_{a, \rm source} \, .
\ee


In our study we distinguish three types of nodes; generators, consumers 
and a slack node. Every node imposes two conditions depending on its type: 
\begin{itemize}
\item
\emph{Generator} nodes have a fixed nominal voltage and provide a fixed 
real power:
\[
  P_a  \stackrel{!}{=}  P_{a, \rm source}, \quad 
  |U_a| \stackrel{!}{=} |U_{a, \rm source}|.
\]
\item
\emph{Consumers} are defined by fixed values of the real and reactive power:
\[
  P_a   \stackrel{!}{=}  P_{a, \rm source}, \quad 
  Q_a  \stackrel{!}{=}  Q_{a, \rm source}  \, .
\]
\item
Furthermore, one introduces a \emph{slack node} which is an ideal voltage source
with the nominal voltage of the grid, i.e. the magnitude and the phase of
the voltage are fixed:
\[
    U_a \stackrel{!}{=}  U_{a, \rm source}.
\]
The slack node compensates any unbalanced 
real or reactive power in the network.
\end{itemize}

In a power grid with $N$ elements, one thus has to solve $2N$ algebraic
nonlinear equations for the $2N$ free variables $\Re(U_a)$, $\Im(U_a)$, $a \in
\{1, \ldots, N\}$. Given the voltages $U_a$ one can easily calculate the power
flows $S_{a,b}$ via Eq.~(\ref{eqn:acflow-power}).

\subsection{Oscillator model}
\label{sup:oscillator model}

We consider a power grid model consisting of $N$ rotating machines 
$j \in \{1,\ldots,N\}$ representing, for instance, wind turbines, or 
electric motors \cite{Prab94,Fila08,12powergrid}. 
Each machine is characterized by the electric power $P_j$ it generates ($P_j > 0$) 
or consumes  ($P_j < 0$). 
The state of each machine is determined by its mechanical phase angle $\theta_j(t)$ 
and its velocity $d \theta_j / dt$. During the regular operation, generators as well 
as consumers within the grid run with the same frequency 
$\omega_0 = 2 \pi \times 50 \, {\rm s}^{-1}$  or 
$\omega_0 = 2 \pi \times 60 \, {\rm s}^{-1}$, respectively. 
The phase of each element is then written as
\begin{equation}
	\theta_j(t) = \omega_0 t + \phi_j(t),
	\label{eqn:phase}
\end{equation}	
where $\phi_j$ denotes the phase difference to the reference phase $\omega_0 t$.

The equation of motion for all $\phi_j$ can now be obtainend from energy 
conservation, that is the generated or consumed energy $P_{\text{source},j}$ of 
each machine must equal the energy sum given or taken from the grid plus 
the accumulated and dissipated energy. The dissipation power of each element 
is given by $P_{\text{diss},j}=\kappa_j (\dot\theta_j)^2$, where $\kappa$ is a
friction coefficient.
The kinetic energy of a rotating machine with a moment of inertia $I_j$ is given 
by $E_{\text{kin},j} = I_j \, \dot \theta^2/2$ such that the accumulated power 
is given by $P_{\text{acc},j} = dE_{\text{kin},j}/dt$.
The power transmitted between two  machines $i$ and $j$ 
is proportional to the sine of the relative phase $\sin(\theta_i - \theta_j)$
and the capacity of the respective transmission line $P_{\text{max},ij}$,
\begin{equation}  
   P_{\text{trans},ij} = P_{\text{max},ij} \sin(\theta_i - \theta_j).
\end{equation}  
If there is no transmission line between two machines, we have $P_{\text{max},ij} = 0$.
The condition of energy conservation at each node $j$ of the network 
now reads
\begin{equation}
   P_{\text{source},j} = P_{\text{diss},j} + P_{\text{acc},j} 
       + \sum_{i=1}^N P_{\text{trans},ij}  \, .
\end{equation}
Note that an energy flow between two elements is only possible if there is a phase 
difference between these two. 

We now insert equation (\ref{eqn:phase}) to obtain the evolution equations for the
phase difference $\phi_j$. We can assume that phase changes are slow compared 
to the set frequency, $|\dot\theta_j| \ll \omega_0$, such that terms containing $\dot \phi_j^2$
and $\dot \phi_j \ddot \phi_j$ can be neglected. Then one obtains 
\begin{equation}
   I_j \omega_0 \ddot\phi_j = P_{\text{source},j}
     -\kappa_j \omega_0^2 - 2  \kappa_j \omega_0 \dot \phi_j + 
     \sum_{i=1}^N P_{\text{max},ij} \sin(\phi_i - \phi_j).
\end{equation}
Note that in the equation only the phase difference $\phi_j$ to the reference phase 
$\omega_0 t$ appears. This shows that only the phase difference between the 
elements of the grid matters. 

For the sake of simplicity we consider similar machines only such that the moment
of inertia $I_j$ and the friction coefficient $\kappa_j$ are the same for all elements
of the network. Defining $P_j := (P_{\text{source},j}-\kappa \omega_0^2)/(I\omega_0)$, 
 $\alpha := 2  \kappa / I$ and $K_{ij} := P_{\text{max},ij} /(I\omega_0)$ this finally leads 
to the equation of motion
\begin{equation}
   \frac{d^2 \phi_j}{dt^2} = P_j - \alpha \frac{d \phi_j}{dt}
          + \sum \nolimits_i K_{ij}\sin(\phi_i-\phi_j) \, .
        \label{eqn:osc-model}  
\end{equation}
Unless stated otherwise, we assume that all transmission lines are equal, that is
\begin{equation}
   K_{ij} = \left\{ \begin{array}{l l}
     K & \; \text{if a link exists between nodes $i$ and $j$} \\
     0 & \text{otherwise.} \\
     \end{array} \right.
\end{equation}


\end{document}